\documentclass[aps,pra,preprint,groupedaddress,showpacs]{revtex4}

\usepackage{graphicx}
\usepackage{bm}
\usepackage{amsmath,amssymb}
\usepackage{txfonts}

\begin{document}


\title{R-matrix calculation of differential cross sections 
for low-energy electron collisions with ground and 
electronically excited state O$_2$ molecules}


\author{Motomichi Tashiro}
\email[]{tashiro@euch4e.chem.emory.edu}
\author{Keiji Morokuma}
\affiliation{Department of Chemistry, Emory University, 1515 Dickey Drive, Atlanta, Georgia 30322, USA.}

\author{Jonathan Tennyson}
\affiliation{Department of Physics and Astronomy,
University College London, London WC1E 6BT, UK.}


\date{\today}

\begin{abstract}
Differential cross sections for electron collisions with the
O$_2$ molecule in its ground ${X}^{3}\Sigma_g^-$ state, as well 
as excited ${a}^{1}\Delta_g$ and ${b}^{1}\Sigma_g^+$ states 
are calculated. As previously, the fixed-bond R-matrix method 
based on state-averaged complete active space SCF 
orbitals is employed. 
In additions to elastic scattering of electron with the 
O$_2$ ${X}^{3}\Sigma_g^-$, ${a}^{1}\Delta_g$ and ${b}^{1}\Sigma_g^+$ 
states, electron impact excitation 
from the ${X}^{3}\Sigma_g^-$ state to the ${a}^{1}\Delta_g$ and 
${b}^{1}\Sigma_g^+$ states as well as '6 eV states' of 
${c}^{1}\Sigma_u^{-}$, ${A'}^{3}\Delta_u$ and ${A}^{3}\Sigma_u^{+}$ 
states is studied. 
Differential cross sections for excitation 
to the '6 eV states' have not been calculated previously.  
Electron impact excitation to the ${b}^{1}\Sigma_g^+$ state 
from the metastable ${a}^{1}\Delta_g$ state is also studied. 
For electron impact excitation from the O$_2$ ${X}^{3}\Sigma_g^-$ state 
to the ${b}^{1}\Sigma_g^+$ state, our results agree better with the  
experimental measurements than previous theoretical calculations. 
Our cross sections show 
angular behaviour similar to  the experimental ones
for transitions from the ${X}^{3}\Sigma_g^-$ state 
to the '6 eV states', although the calculated cross sections are up to 
a factor two larger at large scattering angles. 
For the excitation from the ${a}^{1}\Delta_g$ state to the 
${b}^{1}\Sigma_g^+$ state, our results marginally agree with the
experimental data except for the forward scattering direction.  
\end{abstract}


\maketitle


\section{Introduction}

A detailed knowledge of electron collisions with the oxygen molecule 
is important for the physics and chemistry of both laboratory and astrophysical 
plasmas. 
In particular, recent attempts to understand the electrical discharge 
oxygen-iodine laser have suggested that excited electronic states of O$_2$ molecule 
play an important role \cite{Sh96,Gu04}. 
In a previous paper \cite{Ta06} (henceforth denoted I), 
we studied integral cross sections for electron collisions with 
the O$_2$ molecule in its excited $a{}^{1} \Delta_g$ 
and $b{}^{1} \Sigma_g^{+}$ states, in addition to the much studied 
electron scattering by the O$_2$ ${X}^{3} \Sigma_g^{-}$ ground state. 
We used the fixed-bond R-matrix method with 13 target states represented 
by valence configuration interaction wave functions. 
State-averaged complete active space SCF (SA-CASSCF) orbitals 
with Gaussian type basis functions were employed. 
The calculated cross sections for electron impact excitation from the 
${a}^{1}\Delta_g$ state to the ${b}^{1}\Sigma_g^{+}$ state at 4.5 eV agree
well with the available experimental data of Hall and Trajmar \cite{Ha75}. 
Although elastic scattering of electrons by the ${a}^{1}\Delta_g$ and 
${b}^{1}\Sigma_g^{+}$ states was also studied, we could not find 
any experimental data for comparison. 

In I, theoretical and experimental integral cross sections 
were compared. 
However, differential cross sections (DCSs) provide a more stringent 
test of theory and are often easier to measure reliably
than integral cross sections. 
For electron impact electronic excitations, calculations which 
give good integral cross sections often give DCS's which differ 
significantly from those observed experimentally.
In this paper, we present DCSs for the 
corresponding processes calculated using the same R-matrix model.  

Previous experimental and theoretical studies 
in the field of electron O$_2$ collisions are well summarized by 
Brunger and Buckman \cite{Br02}. 
Here we only review  works relevant to this paper. 
%
The DCSs of electron collisions with 
O$_2$ molecule have been measured by many experimental groups. 
In particular, electron impact excitations to the low-lying 
${a}^{1}\Delta_g$ and ${b}^{1}\Sigma_g^{+}$ states have been 
studied experimentally by Trajmar et al. \cite{Tr71}, 
Shyn and Sweeney \cite{Sh93}, Allan \cite{Al95}, 
Middleton et al. \cite{Mi94}, and Linert et al. \cite{Li04b}.  
In contrast to these experimental works, only  Middleton et al. \cite{Mi94}
report calculations of DCSs for these excitation processes. 
Some of the more recent measurements have focused on electron impact 
excitations from the ${X}^{3} \Sigma_g^{-}$ ground state to 
the `6 eV states', i.e., the ${c}^{1}\Sigma_u^{-}$, ${A'}^{3}\Delta_u$ 
and ${A}^{3}\Sigma_u^{+}$ states which are also called the
Herzberg pseudocontinuum \cite{Ca00,Sh00,Gr02}. 
Although these DCSs are not state-resolved 
in most case, Shyn and Sweeney \cite{Sh00} obtained cross sections 
for excitation to the individual electronic state within the `6 eV states'. 
In this paper, we also calculate the DCSs 
of this process using the fixed-bond R-matrix method, since 
no previous theoretical calculation exists. 
Up to now, there is only one measurement of DCSs 
for electron collisions with electronically excited O$_2$ molecule. 
Hall and Trajmar \cite{Ha75} obtained excitation cross sections from 
the O$_2$ ${a}^{1}\Delta_g$ to the ${b}^{1}\Sigma_g^{+}$ state at 
electron impact energy of 4.5 eV. Their integral cross section was 
compared with our R-matrix results in I. 

In this paper, details of the calculations are presented in section 2, 
and we discuss the results in section 3 comparing our results with 
previous theoretical and available experiments. 
Then the summary is given in section 4. 


\section{Theoretical methods}

The R-matrix method itself has been described extensively in the literature 
\cite{Bu05,Go05,Mo98} as well as in I. 
Thus we do not repeat a general explanation of the method here. 
We used a modified version of the polyatomic programs in the UK molecular 
R-matrix codes \cite{Mo98} to extract T-matrix elements of the 
electron O$_2$ scatterings. 
These programs utilize Gaussian type 
orbitals (GTO) to represent target molecule as well as a scattering electron. 
Although most of the past R-matrix works on electron O$_2$ collisions 
have employed Slater type orbitals (STO), 
we select GTOs mainly because of the simplicity of the input and 
availability of basis functions. 
The SA-CASSCF orbitals are imported from the calculations with 
MOLPRO suites of programs \cite{molpro}. 
The use of SA-CASSCF orbitals improves the vertical 
excitation energies of the O$_2$ target states compared to the 
energies from HF orbitals as shown in I. 
These target orbitals are constructed from the the [5s,3p] contracted basis 
of Dunning\cite{Du71} augmented by a d function with exponent 1.8846, as 
in Sarpal et al. \cite{Sa96}. 
In the R-matrix calculations, we included 13 target states; 
${X}^3\Sigma^{-}_{g}$,${a}^1\Delta_{g}$,
${b}^1\Sigma^{+}_{g}$,${c}^1\Sigma^{-}_{u}$,${A'}^3\Delta_{u}$,
${A}^3\Sigma^{+}_{u}$,${B}^3\Sigma^{-}_{u}$,${1}^1\Delta_{u}$,
${f'}^1\Sigma^{+}_{u}$, 
${1}^1\Pi_{g}$,${1}^3\Pi_{g}$,${1}^1\Pi_{u}$ and ${1}^3\Pi_{u}$, 
where the last 4 $\Pi$ states were not included in the previous
R-matrix studies performed by other groups. 
These target states were represented by valence configuration interaction 
wave functions constructed from the SA-CASSCF orbitals. 
In our fixed-bond R-matrix calculations, these target states are
evaluated at the equilibrium bond length $R$ = 2.3 a$_0$ 
of the O$_2$ ${X}^3\Sigma^{-}_{g}$ ground electronic state. 

The radius of the R-matrix sphere was chosen to be 10 a$_0$ in our  
calculations. 
In order to represent the scattering electron, we included diffuse
Gaussian functions up to $l$ = 5 with 9 functions for $l$ = 0, 7 functions 
for $l$ = 1 - 3 and 6 functions for $l$ = 4 and 5. 
The exponents of these diffuse Gaussian were fitted using the GTOBAS 
program \cite{Fa02} in the UK R-matrix codes.  
Details of the fitting procedure are the same as in Faure 
et al. \cite{Fa02}. 
In addition to these continuum orbitals, we included 8 extra virtual orbitals, one 
for each symmetry. 
The construction of the 17 electrons CSFs for the total system is 
the same as in I. 
The R-matrix calculations were performed for all 8 irreducible 
representations of the D$_{2h}$ symmetry, 
$A_g$, $B_{2u}$, $B_{3u}$, $B_{1g}$, $B_{1u}$, $B_{3g}$, $B_{2g}$ 
and $A_u$, for both doublet and quartet spin multiplicity of the
electron plus target system. 

The DCSs are evaluated from the T-matrix elements 
obtained by the R-matrix calculations. 
As in Gianturco and Jain \cite{Gi86} and Malegat \cite{Ma90}, 
the DCS is expanded using the Legendre polynomials, 
\begin{equation}
\frac{d \sigma}{d \Omega} \bigr|_{ij} = \sum_{k} A_{k} 
P_{k} \left( {\rm cos} \theta \right),
\label{m1eq}
\end{equation}
where $i$ and $j$ denote the initial and final electronic states 
of the target, respectively. 
In exactly the same way as in Malegat \cite{Ma90}, but for $D_{2h}$ 
symmetry instead of $D_{\infty h}$ symmetry in her paper, 
we can derive an expression of the the expansion coefficients 
$A_{k}$, which is 
\begin{eqnarray}
A_{k} &=& \sum_{l_{i} m_{i} l_{j} m_{j} \Gamma \lambda \mu} 
\sum_{l'_{i},m'_{i},l'_{j},m'_{j},\Gamma',\lambda',\mu'} 
\frac{ (-1)^{\mu+\nu} i^{l_{i}-l_{j}-l'_{i}+l'_{j}} \left(2k+1\right) } 
{8\left(2S_{i}+1\right) k_{i}^{2} } \delta_{\lambda'-\lambda,\mu'-\mu} \notag \\
&& \times 
\sqrt{ \left(2l_{i}+1\right) \left(2l'_{i}+1\right) \left(2l_{j}+1\right) \left(2l'_{j}+1\right) } \notag \\  
&& \times 
\begin{pmatrix} l_{i} & l'_{i} & k \\ -\lambda & \lambda' & \lambda-\lambda'
\end{pmatrix}
\begin{pmatrix} l_{j} & l'_{j} & k \\ -\mu & \mu' & \mu-\mu' \end{pmatrix}
\begin{pmatrix} l_{i} & l'_{i} & k \\ 0 & 0 & 0 \end{pmatrix}
\begin{pmatrix} l_{j} & l'_{j} & k \\ 0 & 0 & 0 \end{pmatrix} \notag \\
&& \times 
C_{\lambda,m_{i}} C_{\mu,m_{j}}^{*} C_{\lambda',m'_{i}}^{*} C_{\mu',m'_{j}} 
\sum_{S} \left( 2S+1 \right) T_{il_{i}m_{i},jl_{j}m_{j}}^{\Gamma S M_{S}}
\left( T_{il'_{i}m'_{i},jl'_{j}m'_{j}}^{\Gamma' S M_{S}} \right)^{*}.
\label{m2eq}
\end{eqnarray}
Details of the derivation are given in the Appendix. 
In equation \ref{m2eq},  
$\begin{pmatrix} l_{i} & l'_{i} & k \\ -\lambda & \lambda' & 
\lambda-\lambda' \end{pmatrix}$ etc. are 3$j$ coefficients, $k_i$ is 
the wave number of the incident electron, $S_i$ is the spin quantum number of 
the initial target state, while $S$ is the spin quantum number of the 
total system and $M_S$ is the projection of the total spin.  
The indices $\Gamma$ and $\Gamma'$ run over the 8 irreducible representations 
of the $D_{2h}$ point group, since we employ the polyatomic version of the 
UK R-matrix code. 
The angular quantum numbers of the scattering electron, $l_i$ and $m_i$ etc.
in the T-matrix element $T_{il_{i}m_{i},jl_{j}m_{j}}^{\Gamma S M_{S}}$ 
specify the real spherical harmonics $S_{l}^{m}$ instead of complex form 
$Y_{l}^{m}$, because the $S_{l}^{m}$ transform as irreducible representations 
under $D_{2h}$ symmetry. 
This means, there are transformation matrix elements $C_{\lambda,m_{i}}$ etc. 
in the expression for $A_k$ in order to convert the index of the scattering 
electron from the $S_{l}^{m}$ representation to the $Y_{l}^{m}$ 
representation.  An
expression of the matrix elements $C_{\lambda,m}$ is given in 
the Appendix. 
Finally, we note that the summations over ($\Gamma$,$l_i$,$m_i$) 
should satisfy the symmetry relation, $\Gamma={\rm IR}\left(i\right) \times 
{\rm IR}\left(S_{l_i}^{m_i}\right)$, with IR$\left(i\right)$ and 
IR$\left(S_{l_i}^{m_i}\right)$ each being an irreducible representation 
of the $D_{2h}$ group corresponding to the $i$th target state and the real 
spherical harmonic $S_{l_i}^{m_i}$, respectively. 
This relation also holds for ($\Gamma$,$l_j$,$m_j$), 
($\Gamma'$,$l'_i$,$m'_i$) and ($\Gamma'$,$l'_j$,$m'_j$).

In equation \ref{m2eq}, the T-matrix elements belonging to 
different total symmetries are multiplied together. 
Since these matrix elements come from different calculations, 
overall phases of molecular orbitals and target CI vectors 
underlying these matrix elements may be inconsistent (see \cite{jt195}), 
which may  
result in erroneous relative signs of these T-matrix elements. 
To avoid this inconsistency, we saved reference target CI vectors from 
the first calculation, $A_g$ symmetry for example, and 
then aligned the overall phases of the target 
CI vectors in other calculations, $B_{2u}$, $B_{3u}$, $B_{1g}$, 
$B_{1u}$, $B_{3g}$, $B_{2g}$, $A_u$ symmetries,   
according to this reference.   
The same set of molecular orbitals was used in all these calculations.


\section{Results and discussion}

\subsection{Electron collisions with the O$_2$ ${X}^3\Sigma^{-}_{g}$ ground state}

\begin{figure}
\includegraphics{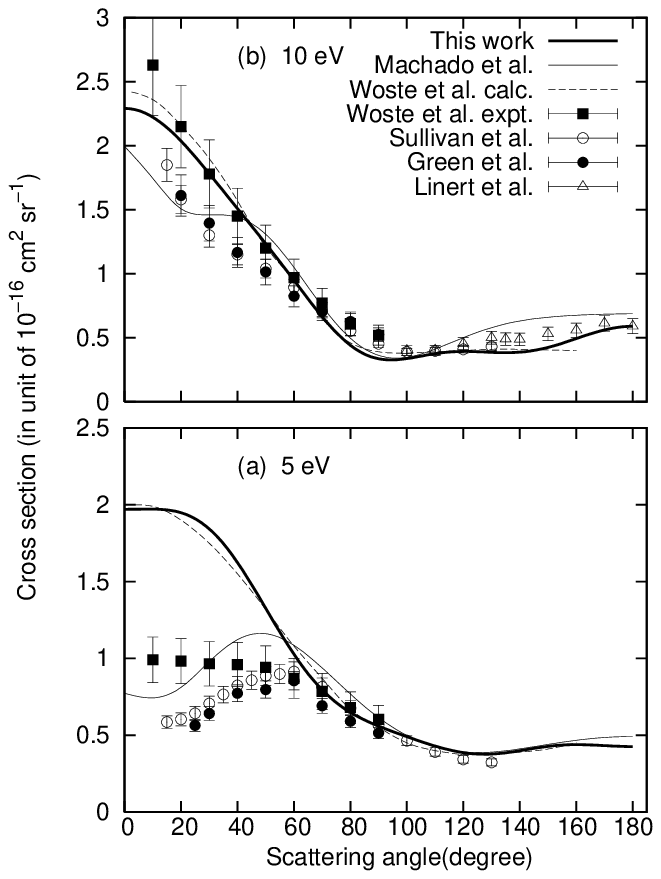}%
 \caption{\label{fig1} 
  Differential cross sections for elastic electron collisions with 
  the O$_2$ ${X}^3\Sigma^{-}_{g}$ state. 
  Panel (a): electron impact energy of 5 eV and (b):10 eV.  
  Thick full line represents our result. 
  For comparison, previous theoretical results of 
  {W\"oste} et al. \cite{Wo95} and Machado et al. \cite{Ma99} are shown as 
  thin lines. Symbols with error bars indicate experimental cross sections 
  of {W\"oste} et al. \cite{Wo95}, Sullivan et al. \cite{Su95}, 
  Green et al. \cite{Gr97} and Linert et al. \cite{Li04}.  
   }
\end{figure}

Figure \ref{fig1} shows DCSs for elastic 
electron scattering from the O$_2$ ${X}^3\Sigma^{-}_{g}$ state 
compared with previous theoretical and experimental results. 
Our results are very similar to the previous R-matrix cross sections 
of {W\"oste} et al. \cite{Wo95}.
The cross sections of Machado et al. \cite{Ma99} 
were calculated using the Schwinger variational iterative method 
combined with the distorted-wave approximation. 
Their results at 5 eV are much lower than the R-matrix results at low 
scattering angle below 50 degrees. 
Our results agree reasonably well with the experimental cross sections 
at 10 eV, including the recent results of 
Linert et al. \cite{Li04} for backward scattering. 
At 5 eV, our model significantly overestimates the cross sections 
for forward scattering compared to the experimental values. 
For example, our result is twice as large as the experimental 
values at 10$^\circ$. 
This situation is the same in the previous R-matrix calculation of 
{W\"oste} et al. \cite{Wo95}. 
As discussed by {W\"oste} et al., this deviation can be attributed to 
a lack of long-range polarizability in the scattering model. 
For example, Gillan et al. \cite{Gi88} introduced polarized pseudostates 
to account for the long-range polarizability in electron-N$_2$ scattering 
and reduced the cross sections by 50\%\ in the threshold energy region. 
The interaction potential of Machado et al. \cite{Ma99} includes the 
correlation-polarization term based on free-electron-gas 
model. Probably the polarization introduced by this term is responsible 
for their better agreement with experiment at 5 eV. 
Since we are interested in electron collisions with the excited electronic 
states of O$_2$ in this work, we chose not to pursue precise accuracy 
further for the ground state elastic scattering. 
However, we have to be mindful that similar long-range polarizability 
problems may exist in the other low-energy electron scattering processes, 
especially elastic electron scattering of the ${a}^1\Delta_{g}$ 
and ${b}^1\Sigma^{+}_{g}$ state O$_2$ molecules,
which will be discussed below. 

\begin{figure}
 \includegraphics{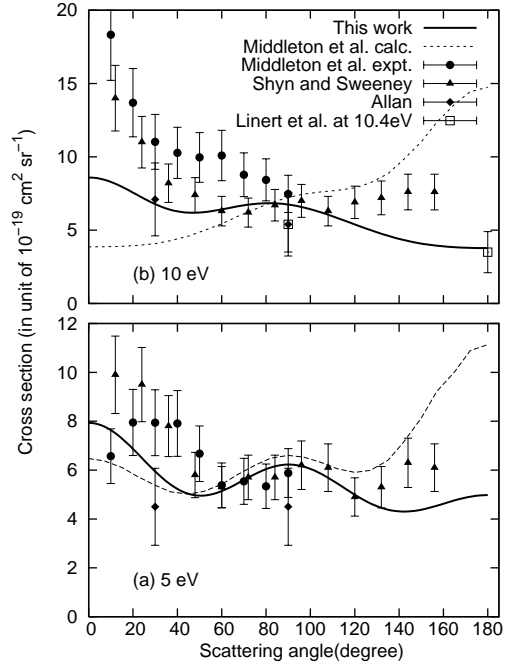}%
 \caption{\label{fig2} 
  Differential cross sections for electron impact excitation from the 
  O$_2$ ${X}^3\Sigma^{-}_{g}$ state to the ${a}^1\Delta_{g}$ state. 
  Panel (a): electron impact energy of 5 eV and (b):10 eV.  
  Full line represents our result. 
  For comparison, we include previous theoretical and experimental 
  cross sections of Middleton et al. \cite{Mi94}, experimental results of   
  Shyn and Sweeney \cite{Sh93}, Allan \cite{Al95} 
  and Linert et al. \cite{Li04b}.
   }
\end{figure}

The DCSs for excitation to the ${a}^1\Delta_{g}$ 
state at electron impact energy 5 and 10 eV 
are compared in figure \ref{fig2} with the 
previous theoretical calculation and the experimental measurements 
of Middleton et al. \cite{Mi94}, Shyn and Sweeney \cite{Sh93}, 
Allan \cite{Al95} and Linert et al. \cite{Li04b}. 
The cross sections at 5 eV agrees well with the previous calculation and 
experimental data below 120$^\circ$. 
However, our results are much lower than the previous 
calculation of Middleton et al. at scattering angle above 130$^\circ$.  
At an electron scattering energy of 10 eV, our cross section deviates 
further from the previous calculation of Middleton et al. \cite{Mi94} 
especially at scattering angle below 60$^\circ$ and above 140$^\circ$. 
In contrast to the backward enhanced cross sections of Middleton et al., 
our DCSs have a slightly forward enhanced character.  
Our results agree better with the experimental data at low scattering angles
than the previous calculation of Middleton et al. \cite{Mi94}. 
At large scattering angle, our results deviate from the experimental 
data of Shyn and Sweeney \cite{Sh93}, but agrees rather well with 
the recent measurement of Linert et al. \cite{Li04b}. 
The R-matrix model of Middleton et al. \cite{Mi94} included the lowest 9 
O$_2$ target states and $l$ = 0 - 5 scattering electron orbitals with 
$\sigma$, $\pi$ and $\delta$ symmetry. In this work, we included 
13 target states and all components of $l$ = 0 - 5 scattering electron 
orbitals. 
In addition to these differences, Middleton et al. used HF/STO orbitals 
where we employed CASSCF/GTO orbitals. 
We carried out a test calculation with  $l$ = 0 - 3 scattering electron 
orbitals and got almost the same cross sections as in $l$ = 0 - 5 case, 
which suggests that difference in the number of target states may be 
important for the shape of these excitation cross sections. 

\begin{figure}
 \includegraphics{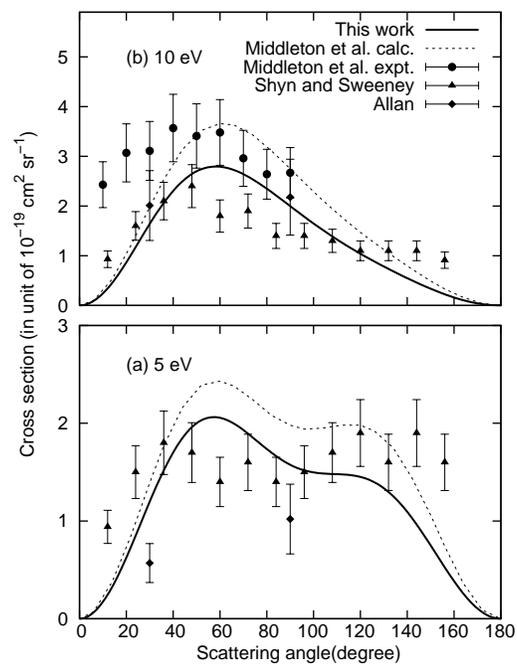}%
 \caption{\label{fig3} 
  Differential cross sections for electron impact excitation from the 
  O$_2$ ${X}^3\Sigma^{-}_{g}$ state to the ${b}^1\Sigma^{+}_{g}$ state. 
  Other details are the same as in figure \ref{fig2}. 
   }
\end{figure}

Figure \ref{fig3} compares DCSs for excitation to 
the ${b}^1\Sigma^{+}_{g}$ state at electron impact energy of 5 and 10 eV 
with the previous R-matrix calculation and the experimental 
measurements of Middleton et al. \cite{Mi94}, Shyn and Sweeney \cite{Sh93} 
and Allan \cite{Al95}. Transitions between $\Sigma^{+}$ and 
$\Sigma^{-}$ target states are forbidden at scattering angles 
of 0$^\circ$ and 180$^\circ$,
because the scattered electron wave function vanishes 
in the plane defined by incident electron beam and the molecular axis 
for any orientation of the molecule \cite{Ca71,Go71}. 
As a consequence, the DCSs decrease to be zero 
toward 0 and 180$^\circ$. 
As is apparent from figure \ref{fig3}, our cross sections become 
zero at 0 and 180$^\circ$, which is consistent with this selection rule. 
Compared to the previous R-matrix calculations of 
Middleton et al. \cite{Mi94}, 
our cross sections have similar profile, but with slightly smaller 
magnitude at all scattering angles. 
Agreement with experiment is good at 5 eV below 120$^\circ$, 
although our results underestimate the experimental cross sections 
at larger scattering angles.  
At 10 eV, the magnitude of the experimental cross sections of Middleton 
et al. \cite{Mi94} and Shyn and Sweeney \cite{Sh93} do not agree with 
each other, however our cross sections are closer to the results of 
Shyn and Sweeney at low scattering angles below 50$^\circ$. 
Between 60$^\circ$ and 90$^\circ$, our results are closer to the results of 
Middleton et al. 

\begin{figure}
 \includegraphics{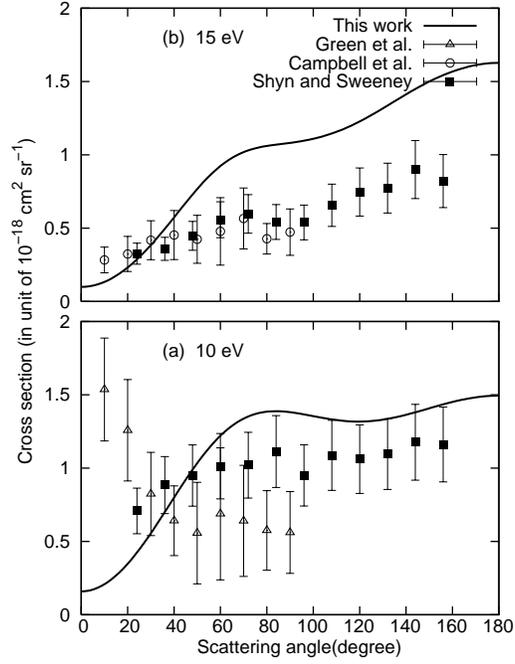}%
 \caption{\label{fig4} 
  Differential cross sections for excitation from 
  the O$_2$ ${X}^3\Sigma^{-}_{g}$ 
  state to the `6 eV states' which consist of the 
  ${c}^1\Sigma^{-}_{u}$, ${A'}^3\Delta_{u}$ and 
  ${A}^3\Sigma^{+}_{u}$ states. The cross sections shown here 
  are sum of the individual cross sections of these 3 states. 
  Panel (a): electron impact energy of 10 eV and (b):15 eV.  
  For comparison, we include experimental results of 
  Green et al. \cite{Gr02}, Campbell et al. \cite{Ca00} and 
  Shyn and Sweeney \cite{Sh00}. 
   }
\end{figure}

\begin{figure}
 \includegraphics{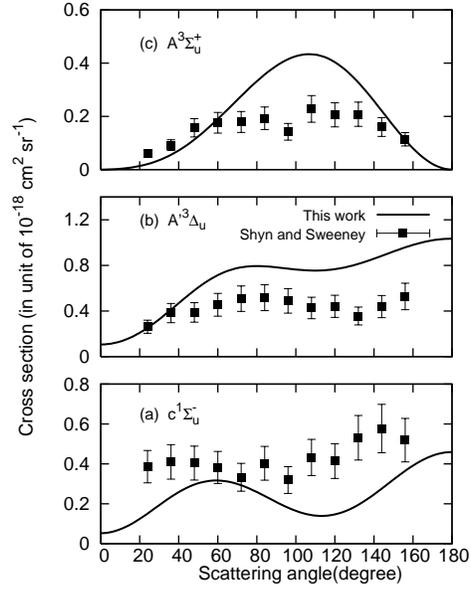}%
 \caption{\label{fig5} 
  Differential cross sections for excitation from 
  the O$_2$ ${X}^3\Sigma^{-}_{g}$ 
  state to the individual state of the `6 eV states'.
  Panel (a) shows excitation cross sections for the ${c}^1\Sigma^{-}_{u}$ 
  state, panel (b) is for the ${A'}^3\Delta_{u}$ state and 
  panel (c) is for the ${A}^3\Sigma^{+}_{u}$ state. 
  The electron impact energy is 10 eV. 
  The experimental data was taken from Shyn and Sweeney \cite{Sh00}. 
   }
\end{figure}

\begin{figure}
 \includegraphics{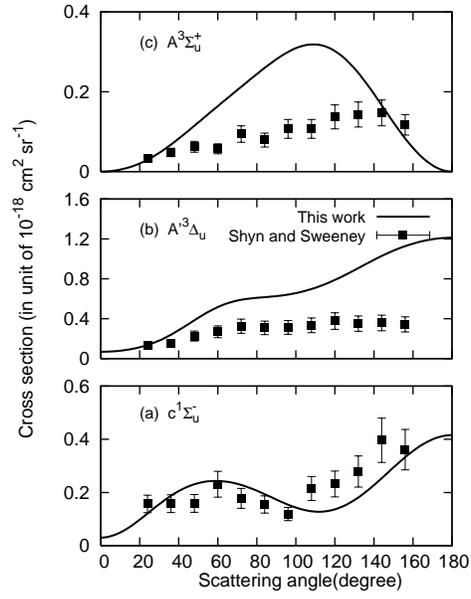}%
 \caption{\label{fig6} 
 The same as in figure \ref{fig5}, but  
 for an electron impact energy of 15 eV. 
   }
\end{figure}

Figure \ref{fig4} shows DCSs for excitations 
to the '6 eV states' for electron impact energies of 
10 and 15 eV. Here the '6 eV states' means the group 
of the O$_2$ ${c}^1\Sigma^{-}_{u}$, ${A'}^3\Delta_{u}$ and 
${A}^3\Sigma^{+}_{u}$ states. The cross sections shown in 
figure \ref{fig4} are a sum of individual excitation cross sections of 
these 3 electronic states, in line with most experimental 
measurements. 
The figure includes the recent experimental cross sections of 
Campbell et al. \cite{Ca00}, Green et al. \cite{Gr02} and 
Shyn and Sweeney \cite{Sh00}. 
The individual cross sections are shown in figure \ref{fig5} and 
\ref{fig6} for impact energies of 10 and 15 eV, together with the
state-resolved experimental cross sections of Shyn and Sweeney 
\cite{Sh00}. 
Our summed cross sections given in figure \ref{fig4} are 
backward-enhanced for both the 10 and 15 eV cases, in accordance with  
the experimental cross sections Campbell et al. \cite{Ca00} and 
Shyn and Sweeney \cite{Sh00}. However, the forward enhancement of the 
DCSs at 10 eV observed by 
Green et al. \cite{Gr02} is not reproduced by our calculation. 
The individual cross sections in figure \ref{fig5} and 
\ref{fig6} show similar angular behaviour compared to the 
experimental results of Shyn and Sweeney. However 
our DCSs for excitation to the ${A'}^3\Delta_{u}$ state is 
more steep toward backward direction.  
Also the peak in the ${A}^3\Sigma^{+}_{u}$ state cross sections 
is more pronounced in our calculation. 
Note that our results for the ${A}^3\Sigma^{+}_{u}$ state become zero 
at 0 and 180$^\circ$ as dictated by the $\Sigma^{-}-\Sigma^{+}$ 
selection rule. 

\subsection{Electron collisions with the O$_2$ ${a}^1\Delta_{g}$ and 
${b}^1\Sigma^{+}_{g}$ excited states }

\begin{figure}
 \includegraphics{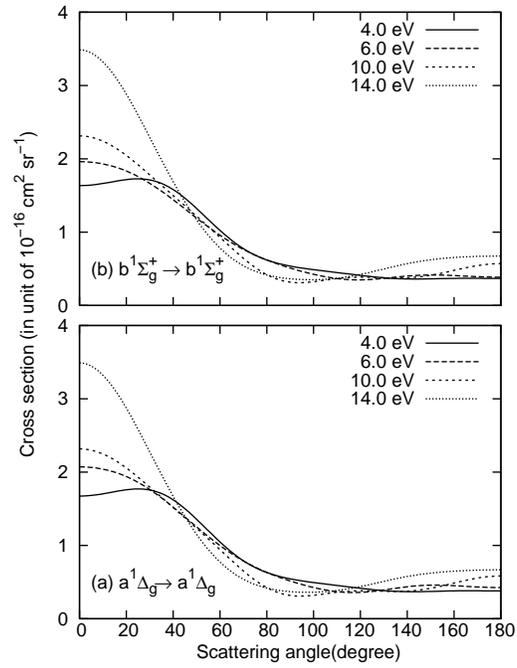}%
 \caption{\label{fig7} 
Differential cross sections for elastic scattering of 
the O$_2$ excited states. (a): the ${a}^1\Delta_{g}$ state,  
(b):  the ${b}^1\Sigma^{+}_{g})$ state. 
Each line corresponds to a different electron impact energy as shown 
in the legend. 
   }
\end{figure}

The DCSs for elastic electron scattering with 
the excited O$_2$ ${a}^1\Delta_{g}$ and ${b}^1\Sigma^{+}_{g}$ states 
are shown in figure \ref{fig7}. 
We cannot compare them with previous theoretical or experimental work, 
since there is no available data. 
These DCSs show strong similarity 
with those of the elastic electron scattering with the 
${X}^3\Sigma^{-}_{g}$ ground state in figure \ref{fig1}. 
The magnitude of these cross sections is almost the same for the 
10 eV case. 
All of them have a large forward peak at 0$^\circ$, a small rise in the cross 
sections at 180$^\circ$. The location of the minimum moves inward 
from 140$^\circ$ to 90$^\circ$ as the electron scattering energy increases. 
This similarity is also reflected in the integral cross sections for 
elastic electron collisions with the ${X}^3\Sigma^{-}_{g}$, 
${a}^1\Delta_{g}$ and ${b}^1\Sigma^{+}_{g}$ states. 
The profiles and magnitudes of the integral cross sections 
are basically the same for all these 3 electronic states as shown in I. 
The main configuration of these 3 electronic states has the form 
$\left( {\rm core} \right) \pi_g^4 \pi_u^2 $, and this may be 
responsible for this similarity. 
Our R-matrix calculations tend to overestimate the elastic scattering 
cross sections of the ${X}^3\Sigma^{-}_{g}$ state at low scattering 
angles, below 50$^\circ$, compared to the experimental data. 
Considering the strong similarity of the cross section profiles 
for elastic scattering from excited states and the ground state, 
our calculations may also overestimate the cross section at low scattering 
angle at low electron impact energy. 

In table \ref{tab1}, we show momentum transfer cross sections for electron 
elastic scattering by the ${X}^3\Sigma^{-}_{g}$, ${a}^1\Delta_{g}$ 
and ${b}^1\Sigma^{+}_{g}$ states. 
As a consequence of the similarity in DCSs, 
the momentum transfer cross sections have a similar magnitude. 
Compared to the experimental data of Shyn and Sharp \cite{Sh82} and 
Sullivan et al. \cite{Su95}, our calculation overestimates the 
${X}^3\Sigma^{-}_{g}$ state momentum transfer cross section 
at 2 eV by 20 \%, but underestimates the cross section at 10 eV 
by 8 \% of Sullivan et al.'s value or 29 \% of Shyn and Sharp's value.  
Our momentum transfer cross sections for the ${a}^1\Delta_{g}$ and 
${b}^1\Sigma^{+}_{g}$ states may similarly be overestimates 
or underestimates depending on the electron impact energy. 

\begin{figure}
 \includegraphics{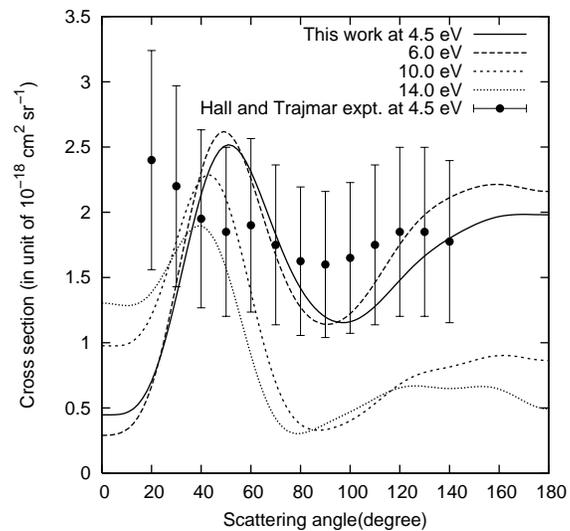}%
 \caption{\label{fig8} 
  Differential cross sections for electron impact excitation from the 
  O$_2$ ${a}^1\Delta_{g}$ state to the ${b}^1\Sigma^{+}_{g}$ state. 
  Each line corresponds to a different electron impact energy, as shown 
  in the legend. 
  Experimental cross sections at 4.5 eV of Hall and Trajmar \cite{Ha75}
  are also shown. 
   }
\end{figure}

Figure \ref{fig8} shows DCSs for  
electron impact excitation from the ${a}^1\Delta_{g}$ state
to the ${b}^1\Sigma^{+}_{g}$ state. 
The figure also includes the experimental data of Hall and Trajmar \cite{Ha75} 
at impact energy of 4.5 eV. 
Our cross section profiles have characteristic features 
of minima around 10$^\circ$ and 90$^\circ$ and maxima around 50$^\circ$ and 150$^\circ$.   
They agree with the experimental cross sections of 
Hall and Trajmar \cite{Ha75} within their error bars except at 20$^\circ$ and 30$^\circ$. 
The cross sections of Hall and Trajmar appear to increase from 50$^\circ$ 
to 0$^\circ$ whereas our cross sections decrease from 60$^\circ$ toward 
10$^\circ$. 
In the 50$^\circ$ -- 140$^\circ$ angular region, Hall and Trajmar's cross sections 
vary less than ours. 
However, a precise comparison is difficult because of large 
error bars and lack of other experimental data. 



\section{Summary}

We have calculated differential cross sections for electron collisions with 
O$_2$ molecule in its ground ${X}^{3}\Sigma_g^-$ state, as well as excited 
${a}^{1}\Delta_g$ and ${b}^{1}\Sigma_g^+$ states. 
As in our previous work, we employed the fixed-bond R-matrix method based 
on state-averaged complete active space SCF orbitals. 
In additions to elastic scattering of electron with the 
O$_2$ ${X}^{3}\Sigma_g^-$, ${a}^{1}\Delta_g$ and ${b}^{1}\Sigma_g^+$ 
states, we studied electron impact excitations from the 
${X}^{3}\Sigma_g^-$ state to the ${a}^{1}\Delta_g$ and ${b}^{1}\Sigma_g^+$ 
states as well as '6 eV states' of ${c}^{1}\Sigma_u^{-}$, 
${A'}^{3}\Delta_u$ and ${A}^{3}\Sigma_u^{+}$ states. 
DCSs for the excitations to the '6 eV states' were 
not calculated previously. 
We also studied electron impact excitation to the ${b}^{1}\Sigma_g^+$ 
state from the metastable ${a}^{1}\Delta_g$ state. 
For electron impact excitation from the O$_2$ ${X}^{3}\Sigma_g^-$ state 
to the ${b}^{1}\Sigma_g^+$ state, our results agree better with the  
experimental measurements than the previous theoretical cross sections. 
Our cross sections show similar angular behaviour the to experimental ones 
for transitions from the ${X}^{3}\Sigma_g^-$ state to the '6 eV states'. 
For the excitation from the ${a}^{1}\Delta_g$ state to the 
${b}^{1}\Sigma_g^+$ state, 
our results marginally agree with experimental data except for the 
forward scattering direction.


\begin{acknowledgments}
The present research is supported in part by the grant from the Air Force 
Office of Scientific Research: the Advanced High-Energy Closed-Cycle Chemical 
Lasers project (PI: Wayne C. Solomon, University of Illinois, 
F49620-02-1-0357). Computer resources were provided in part by the Air Force 
Office of Scientific Research DURIP grant (FA9550-04-1-0321) as well as 
by the Cherry L. Emerson Center for Scientific Computation at Emory University.
The work of M.T. was supported by the Japan Society for the 
Promotion of Science Postdoctoral Fellowships for Research Abroad. 
\end{acknowledgments}

\clearpage


\clearpage


%



\clearpage

\clearpage

%

\clearpage

%

\clearpage

%

\clearpage

\clearpage

\clearpage

\clearpage

%

%

\begin{table}
\caption{\label{tab1}
Elastic momentum transfer cross sections in unit of $10^{-16}$cm$^2$.
}
\begin{ruledtabular}
\begin{tabular}{rrrrrrr}
Electron impact energy(eV)   & 2.0  & 4.0 & 6.0 & 8.0 & 10.0 & 14.0 \\
\hline
This work ~ ~ ~ ~ ~ ~ ${X}^3\Sigma^{-}_{g}$  & 8.18 & 6.80 & 6.73 & 6.06 & 6.04 & 6.96   \\
${a}^1\Delta_{g}$  & 8.16 & 6.69 & 6.57 & 5.90 & 5.86 & 6.79   \\
${b}^1\Sigma^{+}_{g}$  & 7.95 & 6.57 & 6.29 & 5.83 & 5.85 & 6.73  \\
\hline
Shyn and Sharp \cite{Sh82} ~ ~   ${X}^3\Sigma^{-}_{g}$  & 6.7 & -   & - & -   & 8.4 & -   \\
Sullivan et al. \cite{Su95} ~ ~ ${X}^3\Sigma^{-}_{g}$  & 6.5 & 6.0 & - & 6.2 & 6.4 & -  
\end{tabular}
\end{ruledtabular}
\end{table}



\clearpage

\appendix*
\section{Derivation of equation \ref{m2eq}.}

Derivation of the DCS formula in equation \ref{m2eq} is similar to that of
Malegat \cite{Ma90} except for the use of real spherical harmonics 
$S_l^m$ employed in the polyatomic version of UK molecular R-matrix codes 
instead of complex form $Y_l^m$. 
For convenience of reader, brief derivation of the formula is given 
in this appendix. In the expressions below, we follow the notation of 
Malegat \cite{Ma90}. 

The scattering wave function describing collision of an electron 
plane wave with a molecule is expressed as,
\begin{equation}
\Psi_{I} \left(x'_1,..,x'_N,\sigma',\bm{r} \right) = 
\Psi_{i} \left(x'_1,..,x'_N \right) \chi_{\frac{1}{2}m_{s_{i}}} 
\left( \sigma' \right) e^{i k_{i} z} + 
\sum_{J} \Psi_{j} \left(x'_1,..,x'_N \right) \chi_{\frac{1}{2}m_{s_{j}}}
\left( \sigma' \right) 
F_{IJ} \left( \hat{\bm{r}} \right) e^{i k_{j} r} / r . 
\label{a1eq}
\end{equation}
Here $x$ denotes the space and spin coordinates of the molecular electrons. 
The primed coordinates refer to the molecular frame with $z'$-axis 
along the molecular symmetry axis, and the unprimed coordinates to 
the laboratory frame with the $z$-axis along the incident electron beam. 
The incident electron has wavenumber $k_i$ with spin projection $m_{s_i}$.  
The index $i$ represents quantum numbers of the electronic 
state of the target molecule, $\Gamma_i$, $S_i$ and $M_{S_i}$, whereas 
the index $I$ refers to ($i$, $m_{s_i}$) collectively.

In order to expand the wave function in equation \ref{a1eq}, a
symmetry adapted N+1-electron wave function is prepared as
\begin{equation}
\Psi_{\bar{i}l_{i}m_{i}}^{\Gamma S M_{S}} 
\left(x'_1,..,x'_N,\sigma',\bm{r} \right) = \sum_{\bar{j}l_{j}m_{j}}
\Psi_{\bar{j}}^{S M_{S}} \left(x'_1,..,x'_N,\sigma' \right)
S_{l_{j}}^{m_{j}} \left( \hat{\bm{r}'} \right) 
f_{\bar{i}l_{i}m_{i},\bar{j}l_{j}m_{j}}^{\Gamma S M_{S}} \left( r \right) / r, 
\label{a2eq}
\end{equation}
where $\Gamma$, $S$ and $M$ stand for symmetry of the N+1-electron system, 
i.e., an irreducible representation of the $D_{2h}$ group in this work, 
spin quantum number and its projection to the symmetry axis. 
The orbital angular momentum of the scattering 
electron and its projection are represented by $l_i$ and $m_i$.  
In case of $m > 0$, the real spherical harmonics $S_l^m$ is related to 
the complex form of spherical harmonics $Y_l^m$ as \cite{He00}
\begin{equation}
\begin{pmatrix} Y_{l}^{m} \\ Y_{l}^{-m} \end{pmatrix} =
\frac{1}{\sqrt{2}}
\begin{pmatrix} (-1)^{m} & (-1)^{m} i \\ 1 & -i \end{pmatrix}
\begin{pmatrix} S_{l}^{m} \\ S_{l}^{-m} \end{pmatrix}. 
\label{a3eq}
\end{equation}
In the $m=0$ case, we only have $Y_l^0$ and $S_l^0$ and thus 
the matrix element is 1. 
Note that $S_l^m$ behaves as an irreducible representation under $D_{2h}$ 
symmetry operations, whereas $Y_l^m$ does not. 
The spin coupled function in equation \ref{a2eq} is given by
\begin{equation}
\Psi_{\bar{j}}^{S M_{S}} \left(x'_1,..,x'_N,\sigma' \right) =
\sum_{M_{S_{j}}m_{s_{j}}} 
\langle S_{j}M_{S_{j}},\frac{1}{2}m_{s_{j}} | S M_{S} \rangle
\Psi_{j} \left(x'_1,..,x'_N \right) 
\chi_{\frac{1}{2}m_{s_{j}}} \left( \sigma' \right),
\label{a4eq}
\end{equation}
where $\langle S_{j}M_{S_{j}},\frac{1}{2}m_{s_{j}} | S M_{S} \rangle$ 
refers to the Clebsch-Gordan coefficients and 
$\bar{i}$ represents $\Gamma_i$ and $S_i$. 
The radial functions in equation \ref{a2eq} are related to the S-matrix 
in the asymptotic region by
\begin{equation}
\lim_{r\to\infty}
f_{\bar{i}l_{i}m_{i},\bar{j}l_{j}m_{j}}^{\Gamma S M_{S}} \left( r \right) = 
\frac{1}{\sqrt{k_{j}}}
\left[
e^{-i\left( k_{j}r-\frac{1}{2}l_{j}\pi \right)} 
\delta_{\bar{i}l_{i}m_{i},\bar{j}l_{j}m_{j}}
- e^{+i\left( k_j r-\frac{1}{2}l_j \pi \right)}
S_{\bar{i}l_{i}m_{i},\bar{j}l_{j}m_{j}}^{\Gamma S M_{S}}
\right].
\label{a5eq}
\end{equation}

Expanding equation \ref{a1eq} in the symmetry adapted functions of equation 
\ref{a2eq} gives
\begin{equation}
\Psi_{I} \left(x'_1,..,x'_N,\sigma',\bm{r} \right) =
\sum_{\bar{i}l_{i}m_{i}} \sum_{\Gamma S M_{S}}
a_{\bar{i}l_{i}m_{i}}^{\Gamma S M_{S}} 
\Psi_{\bar{i}l_{i}m_{i}}^{\Gamma S M_{S}} 
\left(x'_1,..,x'_N,\sigma',\bm{r} \right). 
\label{a6eq}
\end{equation}
By comparing the ingoing parts on the right and the left hand side, 
we obtain the expansion coefficient
\begin{equation}
a_{\bar{i}l_{i}m_{i}}^{\Gamma S M_{S}} = 
\frac{ -i^{l_{i}} \sqrt{4\pi \left( 2l_{i}+1 \right)} }{2i \sqrt{k_{i}}}
\sum_{\lambda} \mathcal{D}_{0~\lambda}^{l_{i} ~ *} \left( \alpha \beta \gamma \right) C_{\lambda,m_{i}}
\langle S_{i}M_{S_{i}},\frac{1}{2}m_{s_{i}} | S M_{S} \rangle,
\label{a7eq}
\end{equation}
where $\mathcal{D}_{m~ m'}^{l} \left( \alpha \beta \gamma \right)$ 
is the rotation matrix with the Euler angles $\left(\alpha, \beta, \gamma \right)$
representing rotation of the laboratory frame to the molecular frame. 
The matrix element $C_{\lambda,m}$, defined in equation \ref{a3eq}, 
relates the spherical harmonics $Y_l^{\lambda}$ and $S_l^m$.  
The collision amplitude can then be obtained by equating the outgoing parts, 
\begin{eqnarray}
F_{IJ} \left( \hat{r} \right)
&=& \sum_{l_{i} m_{i} l_{j} m_{j}} \sum_{\Gamma S M_{S} \lambda \mu \nu} 
\frac{ \sqrt{\pi \left(2l_{i}+1\right) } }{\sqrt{k_{i}k_{j}}} 
i^{l_{i}-l_{j}+1} 
\langle S_{i}M_{S_{i}},\frac{1}{2}m_{s_{i}} | S M_{S} \rangle
\langle S_{j}M_{S_{j}},\frac{1}{2}m_{s_{j}} | S M_{S} \rangle   \notag \\
&& \times
\mathcal{D}_{0 ~ \lambda}^{l_{i}~ *} \left( \alpha \beta \gamma \right)
\mathcal{D}_{\nu~\mu}^{l_{j}} \left( \alpha \beta \gamma \right)
Y_{l_{j}}^{\nu} \left( \hat{r} \right) C_{\lambda,m_{i}} C_{\mu,m_{j}}^{*}
T_{\bar{i}l_{i}m_{i},\bar{j}l_{j}m_{j}}^{\Gamma S M_{S}}. 
\label{a8eq}
\end{eqnarray}
Here we use the T-matrix elements 
$T_{\bar{i}l_{i}m_{i},\bar{j}l_{j}m_{j}}^{\Gamma S M_{S}}$ instead of the S-matrix. 

By summing over the final states and averaging over 
the initial states and the molecular orientation 
$\left(\alpha, \beta, \gamma \right)$, 
the differential cross section is expressed by the Legendre polynomials 
expansion \ref{m1eq} with expansion coefficients given by equation \ref{m2eq}.

\end{document}